\documentclass[%
 reprint,
superscriptaddress,
showpacs,
preprintnumbers,
nobibnotes,
 amsmath,amssymb,
 aps,
prl
]{revtex4-1}
\usepackage{url} 
\usepackage{amssymb}
\usepackage{amsbsy}
\usepackage{amsmath}
\usepackage{color}
\usepackage{psfrag,epsfig}
\usepackage{xcolor,graphicx}
\graphicspath{{./figures/}{./}}
\usepackage{times}
\usepackage{mathptmx}

\usepackage[mathlines]{lineno}

\begin{document}


\title{Roughness, Inertia, and Diffusion Effects on Anomalous Transport in Rough Channels}

\author{Seonkyoo Yoon}
\affiliation{Department of Earth Sciences, University of Minnesota, Minneapolis, USA}
\author{Peter K. Kang}%
\email[Corresponding author: ]{pkkang@umn.edu}
\affiliation{Department of Earth Sciences, University of Minnesota, Minneapolis, USA}
\affiliation{Saint Anthony Falls Laboratory, University of Minnesota, Minneapolis, USA}

\date{\today}
\begin{abstract}
We study how the complex interplay between roughness, inertia, and diffusion controls the tracer transport in rough channels. We first simulate the flow and tracer transport over wide ranges of channel roughness, Reynolds number ($Re$), and P\'eclet number ($Pe$) observable in nature. $Pe$ exerts a first-order control on first-passage time distributions, and the effect of roughness on the tracer transport becomes evident with the increase in $Re$. The interplay between the roughness and $Re$ causes eddy flows, which intensify or suppress anomalous transport depending on $Pe$. At infinite $Pe$, the late-time scaling follows a universal power-law scaling, which is explained by conducting a scaling analysis. With extensive numerical simulations and stochastic modeling, we show that the roughness, inertia, and diffusion effects are encoded in Lagrangian velocity statistics represented by velocity distribution and correlation. We finally predict the anomalous transport using a stochastic model that considers the Lagrangian velocity statistics.
\end{abstract}

\maketitle

Fluid flow and mass transport in rough channels are ubiquitous phenomena occurring in numerous engineering applications and natural processes including microfluidics, biomedical devices, heat exchangers, and fractured geological media~\citep{stroock2002chaotic, squires2005microfluidics, sackmann2014present, morini2004single, berkowitz2002characterizing}. Since Taylor’s seminal work on solute dispersion in shear flows~\citep{taylor1953dispersion}, the concept of effective dispersion has proven to be very powerful, and many studies have proposed various methods to quantify the effective dispersion in more complex flow fields~\citep{aris1956dispersion,koplik1988molecular,plumb1988dispersion,frankel1989foundations,sarracino2016nonlinear,detwiler2000solute,gelhar1983three,drazer2001tracer,bouquain2012impact,PhysRevLett.61.2925}. Although these effective dispersion approaches are useful in many applications, the asymptotic regime is often not reached, and anomalous behaviors are widely observed ~\citep{bijeljic2006pore,neuman2009perspective,de2013flow,le2011effective,kang2015impact,nissan2018inertial,cushman2000fractional,benson2001fractional,singha2007geoelectrical,berkowitz1997anomalous}. The solute transport in finite-length channel flows at a relatively high P\'eclet number ($Pe$) is a representative example~\citep{giona2009laminar}. 

Recent studies on tracer transport through channel flow systems focused on the effects of eddies: eddies are shown to trap solute particles and cause heavy tailing of tracer breakthrough curves~\citep{boutt2006trapping,cardenas2009effects,bolster2014modeling,wang2014non}. However, the effects of eddies were investigated for a specific channel geometry and over narrow ranges of Reynolds number ($Re$) and $Pe$. In practice, the roughness, inertia, and diffusion effects can vary over wide ranges, leading to complex transport behaviors~\citep{thompson1991effect,lee2015tail,nissan2018inertial,nissan2019Pe}. For example, the trapping mechanism induced by the diffusion of solute particles into eddies should strongly depend on $Pe$. Currently, we do not have a mechanistic understanding of how the complex interplay between channel roughness, $Re$, and $Pe$ controls the anomalous transport in rough channels.

In this study, we elucidate how this complex interplay governs the tracer transport in rough channels by conducting extensive numerical simulations and stochastic modeling. We consider self-affine rough walls, as rough surfaces in nature are often found to be self-affine~\citep{mandelbrot1983fractal,kertesz1993self,ponson2006two}. Self-affine surfaces are scale invariant in that the height of the rough surface at a spatial location $x$ can be described using $z(x) = \lambda^{-H}z(\lambda x)$, where $H$ is the Hurst exponent that characterizes the roughness of the surface. We investigate the effects of roughness on the flow and transport by varying the Hurst exponent ($H$) in the range of $0.7$ -- $0.9$, which is consistent with that observed in nature~\citep{berkowitz2002characterizing,Bouchaud_1990,drazer2004self}. We use the successive random addition algorithm~\citep{voss1988fractals,liu2004corrected} to generate rough surfaces of length $L = 10$ cm. The generated rough surfaces are duplicated and detached to have a constant aperture of $1$ mm.

\begin{figure*}
\includegraphics[height=2in,width=7in]{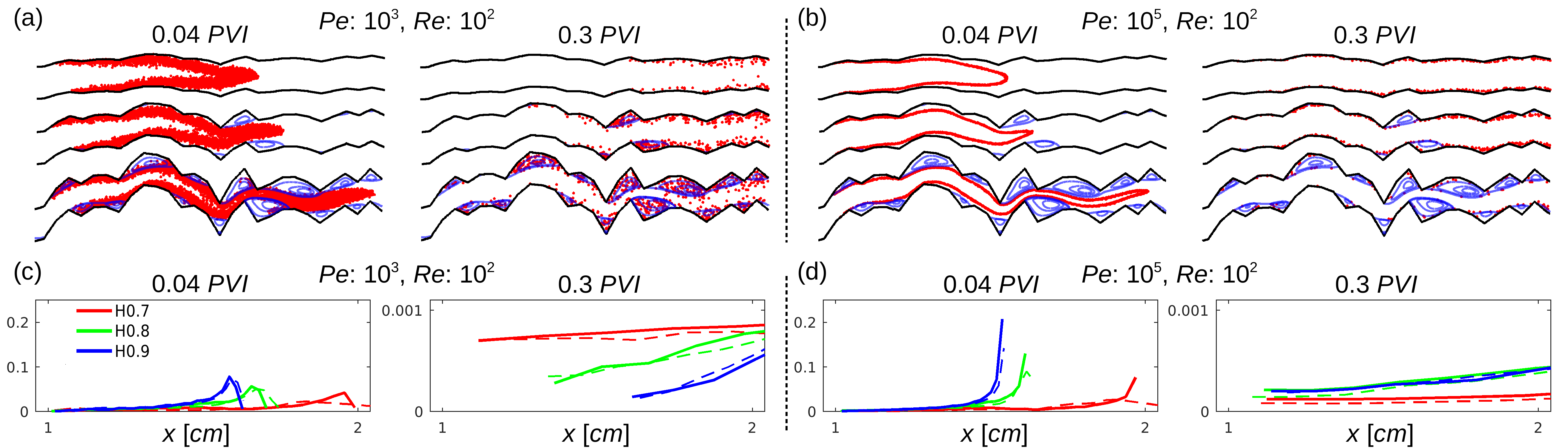}
\caption{\label{fig:geometries}(a, b) Tracer locations between $x=1$ cm and $x=2$ cm at pore volume injections (PVIs) of $0.04$ and $0.3$ are indicated using red circles for $Pe = [10^3, 10^5]$ at $Re=100$, and eddies and streamlines inside the eddies are indicated using blue lines. (c, d) Projected tracer concentration profiles at $0.04$ and $0.3$ PVI for $Pe = [10^3, 10^5]$ at $Re=100$. The dashed lines indicate the CTRW predictions.}
\end{figure*}

 We simulate a Newtonian fluid flow by solving steady-state incompressible Navier--Stokes equations for rough channels using the finite volume method~\citep{OPENFOAM}. A constant flux boundary condition is imposed on the left boundary of the channel, and a zero-pressure gradient boundary condition is imposed on the right boundary. We discretize the channel with a resolution of $0.002$ mm, yielding $50,000\times500$ grid cells within the channel domain. The inertial regimes are quantified with Reynolds number defined as $Re=\frac{ub}{\nu}$, where $b$ is the aperture, and $\nu$ is the kinematic viscosity of the fluid. We consider various laminar flow regimes at seven different Reynolds numbers: $Re=[1,10,20,40,60,80,100]$.

We simulate solute transport using a particle tracking method~\citep{bijeljic2011signature}. In this method, the advective transport is simulated using a streamline-based particle tracking algorithm that considers no-slip boundary conditions at solid--fluid interfaces~\citep{mostaghimi2012}. The diffusive displacement is modeled using a random walk method. The Lagrangian approach is free of numerical dispersion and can be used to accurately simulate particle transport at high $Pe$ regimes. We inject $10^4$ particles in each realization with a flux-weighted line injection, and the $x$-direnctional travel distance is $8$ cm. We investigate the effects of diffusion on the transport by varying the P\'eclet number defined as $Pe=\frac{ub}{2D}$, where $D$ is the molecular diffusivity. We consider five different $Pe$ regimes: $Pe=[10^2, 10^3, 10^4, 10^5, \infty]$. We choose $Pe$ and $Re$ ranges such that they cover the observable Schmidt number, $Sc = \frac{Pe}{Re}$, in nature~\citep{stroock2002chaotic,delgado2006critical}.
In summary, we simulate the flow and transport by varying $H$, $Re$, and $Pe$ independently over wide ranges: $H=[0.7, 0.8, 0.9]$, $Re=[1, 10, 20, 40, 60, 80, 100]$, and $Pe=[10^2, 10^3, 10^4, 10^5, \infty]$. The independent variation in $H$, $Re$, and $Pe$ helps discern the role of each factor. We simulate an ensemble of $10$ realizations for each combination of $H$, $Re$, and $Pe$. 

\paragraph{\label{sec:results} Simulation Results.}
To highlight the importance of understanding the complex interplay, we first present the effects of $H$ and $Pe$ on the tracer transport at $Re = 100$. As shown in Figs.~\ref{fig:geometries}(a) and (b), an increase in the roughness increases the size and frequency of the eddies. The role of eddies on the tracer transport is sensitive to $Pe$. At $Pe=10^3$, tracers readily diffuse into eddies, and the eddies trap these tracers. Interestingly, the eddies play the opposite role at $Pe=10^5$. At $Pe=10^5$, the tracers can no longer easily enter the eddy zones, and the eddies instead act as slip boundaries that help in transporting the tracers near the channel walls (see supporting videos). This is clearly observed in the projected concentration profiles at a pore volume injection (PVI) of $0.3$ (Figs. \ref{fig:geometries}(c) and (d)). The tailing in the profile becomes stronger with the increase in the roughness at $Pe=10^3$; however, the trend is opposite at $Pe=10^5$.

\begin{figure*}
\includegraphics[height=2.4in,width=7in]{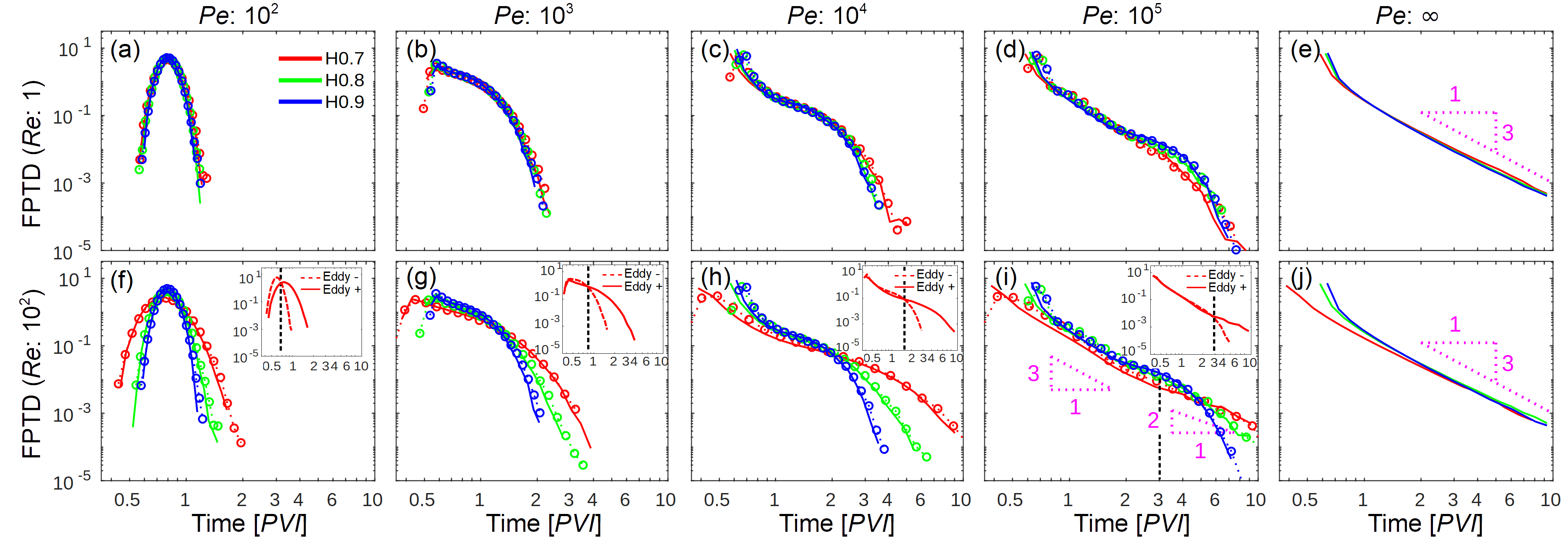}
\caption{\label{fig:FPTD} First-passage time distributions (FPTDs) at $x = 9$ cm. The presented FPTDs are ensemble averages over $10$ realizations. The solid lines indicate the FPTDs from direct simulations, and circles with dashed lines indicate the FPTDs quantified using the CTRW. At $Pe=\infty$, the FPTDs show a universal scaling of $1/3$. Insets: comparison between the FPTD at $H = 0.7$ (Eddy $+$) and the residence time of the tracers in the main channels (Eddy $-$)}
\end{figure*}

The normalized breakthrough curves or first-passage time distributions (FPTDs) for the combinations of $H=[0.7, 0.8, 0.9]$, $Re=[1, 100]$, and $Pe=[10^2,10^3,10^4,10^5, \infty]$ are shown in Fig. \ref{fig:FPTD}. We show FPTDs up to $10$ PVI, though more than $90\%$ of the particles breakthrough within $1$ PVI. First, note that $Pe$ has a first-order control on the breakthrough curve shapes. The effects of roughness only become evident at higher $Re$. We focus on two key characteristics of the anomalous transport: early arrival and late-time tailing in the breakthrough curves. We observe an enhanced early arrival with the increase in the roughness and $Re$. A high roughness coupled with significant inertial effects (high $Re$) causes eddies to enlarge. The eddies narrow the mobile zone, or the main flow channel, and the decreased cross-sectional area of the main flow channel leads to an increase in the velocity along the main flow channel~\citep{zhou2019mass}. Fig. \ref{fig:geometries} shows this phenomenon. This explains the anomalously early arrival of the tracers, shown in Fig. \ref{fig:FPTD}, at high roughness and high $Re$.

The late-time behavior of the FPTDs is determined by the particles traveling through low-velocity regions. No-slip boundary conditions and eddies create low-velocity regions near the channel walls. To analyze the eddy trapping effects, we first delineate eddy zones based on the mass balance principle \citep{zhou2019mass} and quantify the effects of eddies on the first passage time by subtracting the total time spent in the eddies from the first passage time. The subtracted residence time distributions at $H = 0.7$ and $Re = 100$ clearly demonstrate eddy trapping effects (insets in Figs. \ref{fig:FPTD}(f)--(i)). However, with an increase in $Pe$, the trapping effects due to the eddies are observed much later. At $Pe=10^5$, the eddies appear to not delay but rather aid the transit of the particles until $\sim 3$ PVI. This is because the particles cannot easily enter the eddies due to the limited diffusion, and the eddies act as slip boundaries. Therefore, most of the tracers (approximately $99 \%$) breakthrough before $3$ PVI.  Consequently, the FPTDs in the smoother cases (e.g., $H = [0.8, 0.9]$) show stronger tailing than that in the $H = 0.7$ case until $t \sim 3$ PVI (Fig. \ref{fig:FPTD}(i)). We observe that although fewer particles are captured at high $Pe$ regimes, once captured, the trapped particles stay longer inside the eddies compared with that observed at lower $Pe$ regimes. This explains the change in scaling at PVI $\sim 3$ in Fig. \ref{fig:FPTD}(i).

Finally, at $Pe=\infty$, the late-time scaling of the FPTDs shows a universal power law scaling of $t^{-3}$ for all combinations of $Pe$, $Re$, and $H$ (Figs. \ref{fig:FPTD}(e) and (j)). We perform a scaling analysis to explain the observed universal power-law scaling. The low-velocity regions should determine the late-time scaling, and the tracers at $Pe=\infty$ cannot enter the eddies. Thus, the low-velocity distribution at $Pe=\infty$ should be determined by the no-slip boundary conditions. For a Poiseuille flow with an aperture $b$, the velocity profile across the channel follows the parabolic equation $u(y) = 6qb^{-3}\big[(b/2)^2-y^2\big]$, where $q$ is the constant influx into the channel. The Eulerian velocity probability density function (PDF) corresponding to the parabolic profile is given as follows.
\begin{equation}
	f_e(u)=-\frac{2}{b}\frac{dy}{du}=\frac{1}{6q\sqrt{\frac{1}{4b^2}-\frac{u}{6qb}}}.
\end{equation}
The PDF of the Lagrangian velocities is related to the Eulerian velocity PDF through flux weighting as $f_\mathcal {L}(u)=\frac{uf_e(u)}{\int du\,uf_e(u)}$ \citep{dentz2016continuous}. The late-time scaling of the FPTD, $f_t(t)$, should be determined by the distribution of the low velocities in $f_\mathcal {L}(u)$. From $t \propto L/u$ and $f_\mathcal {L}(u)$, the late-time scaling of the FPTD follows,
\begin{equation}
	f_t(t) \propto \frac{1}{t^3} \cdot  \frac{1}{\sqrt{c_1-c_2/t}},
\end{equation}
where $c_1=(4b^2)^{-1}$ and $c_2=(6qb)^{-1}$. For large $t$, we obtain $f_t(t) \propto t^{-3}$. This confirms that the no-slip boundary condition indeed governs the late-time scaling at infinite $Pe$ regardless of the roughness and $Re$.

\paragraph{\label{sec:Vstat} Lagrangian Velocity Statistics.}
Recent studies reported that the pore structure governs the velocity PDF and that pore-scale velocities are strongly correlated~\citep{bijeljic2011signature,de2013flow,datta2013spatial,kang2014pore,de2017prediction,carrel2018pore,crevacore2016recirculation}. The transport in porous media has been successfully characterized by considering the interplay between the velocity PDF and the velocity correlation~\citep{le2008lagrangian,dentz2010distribution,kang2011spatial,bolster2014modeling,kang2015impact,dentz2016continuous}. For channel flows, the roughness of the wall can lead to a significant difference in velocity between the main channel flow and the near-wall low-velocity zones~\citep{meheust2000flow,cardenas2007navier,lee2015tail}. Moreover, the Lagrangian velocities sampled along a particle trajectory in the channel flows should retain the memory of the prior velocities because of the mass conservation constraint. 

We characterize the Lagrangian velocity statistics using the Lagrangian velocity PDF (velocity distribution) and velocity correlation. We first quantify the motion of the solute particles in equidistance, $\Delta x$, in the mean flow direction. The velocity distribution is characterized using the PDF of the transition time $\tau=\frac{\Delta x}{v}$, where $v$ is the average Lagrangian velocity over $\Delta x$. The transition times are sampled at every $\Delta x = 1$ mm from all the particle trajectories. Herein, we refer to the PDF of the transition times as the transition time distribution (TTD). We characterize the Lagrangian velocity correlation by quantifying the velocity correlation lengths conditional to initial velocity values. At preasymptotic regimes, velocity correlation can strongly depend on the initial velocity values \citep{le2007characterization,kang2019}. We classify the initial Lagrangian velocities into $10$ classes, wherein each class is equidistantly spaced in a $\log$-scale. Based on the initial $\log$-velocity values, the particles are assigned into one of the $10$ classes, $i=[1,\dots,10]$, where $i=10$ is the class with the highest velocities. We estimate the characteristic correlation length for each class $i$ as ${\ell}_{i}=\int_0^{L} C(x|i)/C(0|i) dx$, where $C(x|i) = \int_{-\infty}^{\infty} \big| P(\log v | i, x) - P(\log v, L) \big|d\log v$~\citep{le2007characterization}. Here, $P(\log v|i,x)$ is the conditional $\log$-velocity distribution for particles belonging to class $i$, and $P(\log v, L)$ is the marginal $\log$-velocity distribution at the outlet.

\begin{figure}
\includegraphics[height=2.3in,width=3.1in]{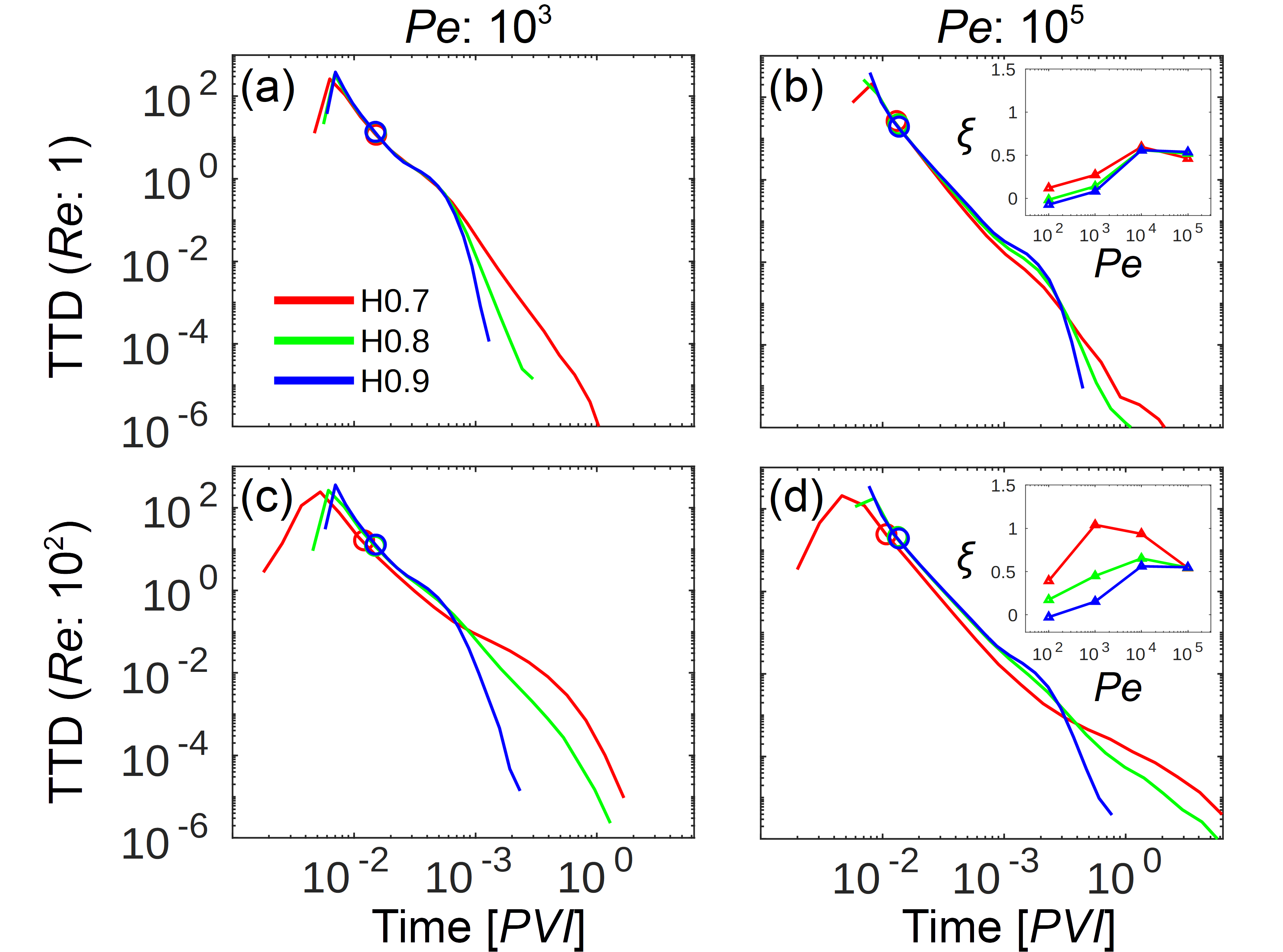}
\caption{\label{fig:heterogneity} Transition time distributions (TTDs). The $0.9$ quantiles are indicated using circles. Insets: tail index $\xi$ as a function of $Pe$ and $H$ at (b) $Re = 1$ and (d) $Re = 100$. $\xi$ quantifies the heavy-tailedness. 
}
\end{figure}

\begin{figure}
\includegraphics[height=2.47in,width=3.2in]{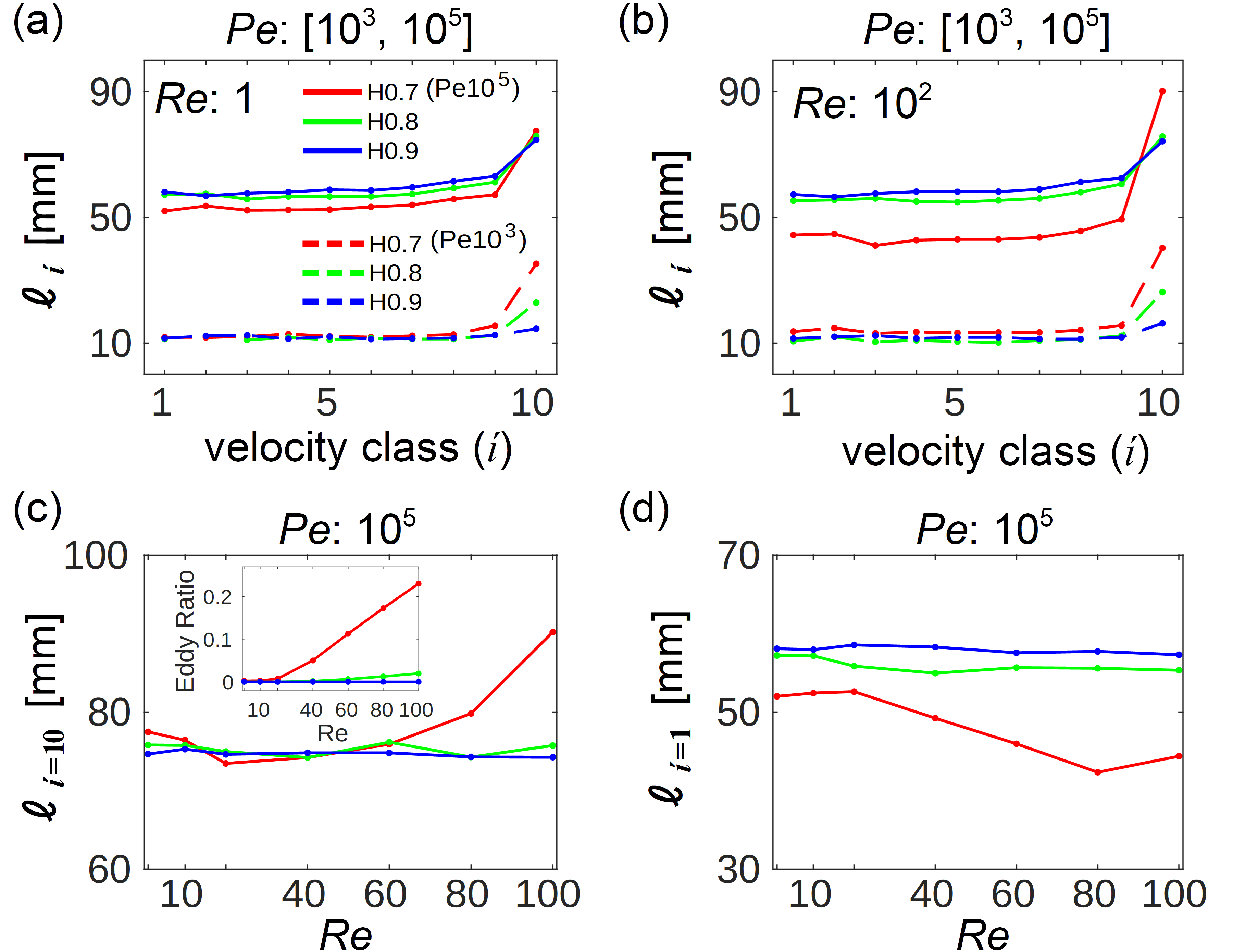}
\caption{\label{fig:correlation} Conditional velocity correlation lengths for $Pe = [10^3, 10^5]$ at (a) $Re = 1$ and (b) $Re = 100$. Correlation length increases with the increase in $Pe$ and velocity class. (c) Evolution of $\ell_i$ as a function of $Re$ at $Pe = 10^5$ for the velocity class $i=10$ and (d) $i=1$. For $H = 0.7$, $\ell_{i=10}$ increases as $Re$ increases, and $\ell_{i=1}$ decreases as $Re$ increases. This behavior is well correlated with the increase in the relative area of the eddy zones (inset of (c)).}
\end{figure}

The underlying mechanisms of the early arrival and late-time tailing are accurately captured in the TTDs (Fig. \ref{fig:heterogneity}) and velocity correlation (Fig. \ref{fig:correlation}). The probability of having short transition times (high-velocity) increases with the increase in the roughness and $Re$ (Fig. \ref{fig:heterogneity}). This implies that the flow channeling effect is encoded in the TTDs. The channeling effect is also encoded in the correlation length (Fig. \ref{fig:correlation}). The correlation length of the velocity class $i = 10$ controls the early arrival of the tracers, and we observe that, for $i = 10$, the correlation length in the $H = 0.7$ case is higher than those in the smoother cases as $Re$ increases (Fig. \ref{fig:correlation}(c)). Eddies, which enlarge with the increase in $Re$ (inset of Fig. \ref{fig:correlation}(c)), reduce the cross-sectional flow area and cause a strong preferential path (Fig. \ref{fig:geometries}). The strong preferential flow reduces the velocity fluctuation and thereby increases the velocity correlation. The higher probability of having higher velocities combined with a strong velocity correlation explains the early arrival behavior at $H = 0.7$.

Similar to the early arrival behavior, the late-time behavior is also encoded in the TTDs and velocity correlation. To quantify the tail of the TTDs, we fit the tail to a generalized Pareto distribution $G_{\xi, \sigma, \theta}(\tau)=\frac{1}{\sigma}(1+ \frac{\xi(\tau-\theta)}{\sigma})^{-1- \frac{1}{\xi}}$ for $\tau>\theta$ when $\xi>0$ or for $\theta<\tau<\theta-\sigma/\xi$ when $\xi<0$. The three parameters $\xi$, $\sigma$, and $\theta$ are estimated using the maximum likelihood method. $\xi$, which is often called the tail index, gives an indication of the heaviness of a tail (the greater the value of $\xi$, the heavier the tail) \citep{coles2001introduction}. The threshold $\theta$ is set as the $0.9$ quantile of the cumulative distribution of the transition times, $\tau_{0.9}$. The insets in Fig. \ref{fig:heterogneity} show the estimates of the tail index. The method of maximum likelihood finds the parameter values that maximize the likelihood function  $\mathcal {L}(\xi, \sigma, \theta \, |\mathbf{t})$ where $\mathbf{t}=\{\tau\,|\tau\ge\tau_{0.9} \}$. 

When eddy flows are strong ($H=0.7$, $Re=100$), the tail index significantly increases with the increase in $Pe$ from $100$ to $10^3$ (inset of Fig. \ref{fig:heterogneity}(d)). This is due to the stronger   eddy trapping effect at $Pe = 10^3$. At $Pe=100$, the particles easily enter the eddies but also easily exit. When $Pe$ increases beyond $10^3$, the tail index decreases significantly at $H=0.7$, confirming that the particles cannot efficiently sample velocities in the eddy zones. Moreover, at high $Pe$ regimes, the particles traveling near the walls will alternate between low (no-slip wall) and relatively high (eddy interface) velocities, thus decreasing the velocity correlation. For the velocity class $i = 1$, the correlation length indeed decreases with the increase in $Re$ at $Pe=10^5$ and $H = 0.7$ (Fig. \ref{fig:correlation}(d)). This trend is well correlated with the increase in the relative eddy area as a function of $Re$ (inset of Fig. \ref{fig:correlation}(c)). The decreased tailing in the TTDs and the loss of velocity correlation for the velocity class $i = 1$ at high $Pe$, high roughness regimes explain how eddies suppress the anomalous transport. Finally, the role of $Pe$ in determining the overall shape of the FPTD is also evident from the TTDs and velocity correlation. $Pe$ has a strong effect on both TTD and velocity correlation, whereas the roughness at $Re = 1$ has little effect on the TTD (inset of Fig. \ref{fig:heterogneity}(b)) and velocity correlation (Fig. \ref{fig:correlation}(a)).

\paragraph{\label{sec:Model} Upscaled Stochastic Transport Model.}
We hypothesize that the complex $H$\textendash$Re$\textendash$Pe$ interplay in rough channels is encoded in the velocity distribution and correlation, which in turn determine the effective transport. To test our hypothesis, we quantify the effective particle transport using an upscaled model that only takes the velocity distribution and correlation as input parameters. The continuous time random walk model with one-step correlation, often referred to as the spatial Markov model, has been successfully applied to predict anomalous transport across scales~\citep{le2008lagrangian,kang2011spatial,de2013flow,kang2014pore,kang2015impact}.

We test the hypothesis by running the spatial Markov model as an effective transport model. The effective particle transport can be characterized using the Langevin equations,
\begin{equation}
	x^{(n+1)}=x^{(n)}+\Delta x, \ \ \
	t^{(n+1)}=t^{(n)}+\frac{\Delta x}{v^{(n)}},
\label{eq:langevin}
\end{equation}
where $\{\tau^{(n)}\}_{n=0}^{L/\Delta x}$ is a series of transition times. The transition time, $\tau^{(n)}$, in Eq. \ref{eq:langevin} is modeled as a Markov chain, whose transitions can be characterized using a transition matrix. We sample the Lagrangian velocity transitions at every $\Delta x = 5$ mm, which is smaller than the estimated correlation length (Fig. \ref{fig:correlation}). We then construct the $10 \times 10$ transition matrix with the $10$ classes equidistantly spaced in a $\log$-scale. The spatial Markov model accurately predicts the projected concentrations and FPTDs for all combinations of $H$, $Re$, and $Pe$ (Figs. \ref{fig:geometries}(c, d) and Fig. \ref{fig:FPTD}). This supports the hypothesis that the $H$\textendash$Re$\textendash$Pe$ interplay is effectively encoded in the velocity distribution and correlation and that they are sufficient for quantifying the effective transport.

In conclusion, we successfully established a mechanistic understanding of how the complex $H$\textendash$Re$\textendash$Pe$ interplay determines the effective tracer transport in rough channel flows. Wide ranges of $H$, $Re$, and $Pe$ observable in nature were investigated, and we found that eddy flows can either induce or nonintuitively suppress the anomalous transport depending on the $Pe$ value. Based on the improved understanding, we proposed a predictive model. For $Pe=\infty$ regime, we observed the universal power-law scaling, $t^{-3}$, in FPTDs and explained it with a scaling analysis.

\subsection{Acknowledgments}
The authors gratefully acknowledge support from NSF via a grant EAR1813526. PKK also acknowledges the College of Science \& Engineering at the University of Minnesota and the George and Orpha Gibson Endowment for its generous support of Hydrogeology. We thank the Minnesota Supercomputing Institute (MSI) at the University of Minnesota for computational resources and support.

\bibliographystyle{apsrev4-1}

%

\end{document}